\documentclass[prb,aps,showpacs,twocolumn,superscriptaddress,nobibnotes,epsf]{revtex4}

\usepackage{graphicx}
\usepackage{dcolumn}
\usepackage{bm}
\usepackage{SIunits}

\begin{document}

\title{Evolution of Anisotropic In-plane Resistivity with doping level in Ca$_{1-x}$Na$_x$Fe$_2$As$_2$ Single Crystals}

\author{J. Q. Ma}
\affiliation{Hefei National Laboratory for Physical Science at
Microscale and Department of Physics, University of Science and
Technology of China, Hefei, Anhui 230026, People's Republic of
China}

\author{X. G. Luo}
\altaffiliation{Corresponding author} \email{xgluo@ustc.edu.cn}
\affiliation{Hefei National Laboratory for Physical Science at
Microscale and Department of Physics, University of Science and
Technology of China, Hefei, Anhui 230026, People's Republic of
China}
\affiliation{Synergetic Innovation Center of Quantum Information \& Quantum Physics, University of Science and Technology of China, Hefei, Anhui 230026, China}

\author{P. Cheng}
\affiliation{Hefei National Laboratory for Physical Science at
Microscale and Department of Physics, University of Science and
Technology of China, Hefei, Anhui 230026, People's Republic of
China}

\author{N. Zhu}
\affiliation{Hefei National Laboratory for Physical Science at
Microscale and Department of Physics, University of Science and
Technology of China, Hefei, Anhui 230026, People's Republic of
China}

\author{D. Y. Liu}
\affiliation{Key Laboratory of Materials Physics, Institute
of Solid State Physics, Chinese Academy of Sciences, P. O. Box 1129,
Hefei 230031, People's Republic of
China}

\author{F. Chen}
\affiliation{Hefei National Laboratory for Physical Science at
Microscale and Department of Physics, University of Science and
Technology of China, Hefei, Anhui 230026, People's Republic of
China}
\affiliation{Synergetic Innovation Center of Quantum Information \& Quantum Physics, University of Science and Technology of China, Hefei, Anhui 230026, China}

\author{J. J. Ying}
\affiliation{Hefei National Laboratory for Physical Science at
Microscale and Department of Physics, University of Science and
Technology of China, Hefei, Anhui 230026, People's Republic of
China}

\author{A. F. Wang}
\affiliation{Hefei National Laboratory for Physical Science at
Microscale and Department of Physics, University of Science and
Technology of China, Hefei, Anhui 230026, People's Republic of
China}

\author{X. F. Lu}
\affiliation{Hefei National Laboratory for Physical Science at
Microscale and Department of Physics, University of Science and
Technology of China, Hefei, Anhui 230026, People's Republic of
China}

\author{B. Lei}
\affiliation{Hefei National Laboratory for Physical Science at
Microscale and Department of Physics, University of Science and
Technology of China, Hefei, Anhui 230026, People's Republic of
China}

\author{X. H. Chen}
\altaffiliation{Corresponding author} \email{chenxh@ustc.edu.cn}
\affiliation{Hefei National Laboratory for Physical Science at
Microscale and Department of Physics, University of Science and
Technology of China, Hefei, Anhui 230026, People's Republic of
China}

\begin{abstract}
We measured the in-plane resistivity anisotropy in the underdoped
Ca$_{1-x}$Na$_x$Fe$_2$As$_2$ single crystals. The anisotropy
(indicated by  $\rho_{\rm b} - \rho_{\rm a}$) appears below a
temperature well above magnetic transition temperature $T_{\rm N}$,
being positive ($\rho_{\rm b} - \rho_{\rm a} > 0$) as $x\leq$ 0.14.
With increasing the doping level to $x$ = 0.19,  an intersection
between $\rho_{\rm b}$ and $\rho_{\rm a}$ is observed upon cooling,
with $\rho_{\rm b} - \rho_{\rm a} < 0$ at low-temperature deep
inside a magnetically ordered state, while $\rho_{\rm b} - \rho_{\rm
a}> 0$ at high temperature. Subsequently, further increase of hole
concentration leads to a negative anisotropy $\rho_{\rm b} -
\rho_{\rm a} < 0$ in the whole temperature range. These results
manifest that the anisotropic behavior of resistivity in the
magnetically ordered state depends strongly on the competition of
the contributions from different mechanisms, and the competition between the
two contributions results in a complicated evolution of the
anisotropy of in-plane resistivity with doping level.
\end{abstract}



\vskip 300 pt

\maketitle

\section{INTRODUCTION}

The undoped and underdoped iron-pnictides undergo structural
transition upon cooling, accompanied with a magnetic transition
from high-temperature paramagnetic to low-temperature
antiferromagnetic (AFM) phase\cite{7}. Understanding the origin of
superconductivity in iron-pnictide might start from the normal-state
physics, especially the roles of the various degrees of freedom in
the magnetic state. One central topic concerning these is the
electronic anisotropy at low temperature, which involves the
fluctuation or ordering of spin, orbital, and band
structures\cite{8,9,10}. In such an anisotropic electronic state,
the striking behavior is that the resistivity along the
ferromagnetically ordered and shorter \emph{b} axis is larger than that
along the antiferromagnetically ordered and longer \emph{a} axis
($\rho_{\rm b} - \rho_{\rm a} > 0$). The anisotropy actually appears
well above the structural and magnetic transition temperatures
$T_{\rm s}$ and $T_{\rm N}$ in electron-underdoped Ba-122
\cite{8,11,12}, which has been discussed according to the nematicity
\cite{25,34,35} by considering the anisotropic magnetic scattering
induced by nematic (spin or/and orbital) fluctuation . The
observation of an orbital ordered polarization of $d_{xz}$ and
$d_{yx}$ of Fe in angle-resolved photoemission spectroscopy (ARPES)
\cite{13}, local anisotropies in scanning tunneling spectroscopy
\cite{14,15}, anisotropies in the optical spectrum \cite{16},
and magnetic susceptibility \cite{31} seems to support the point of
view of nematicity.

However, the origin for the anisotropic in-plane resistivity in the
AFM state remains in hot debate. In undoped BaFe$_2$As$_2$, the
anisotropy has been discussed \cite{9,17} in terms of the
high-mobility Dirac pockets near the Fermi energy, detected by
quantum oscillations and ARPES measurements \cite{18,19,20,21}.
Orbital ordering was also considered as a possible mechanism for the
anisotropy, but calculations based on five-orbital model gave rise
to a sign opposite to that observed in experiments \cite{26,36}.
Another scenario was proposed based on impurity scattering to
interpret such anisotropic in-plane resistivity \cite{12,22,23}.
Especially, annealing can lead to almost annihilation of transport
anisotropy at low temperature in undoped BaFe$_2$As$_2$
\cite{22,23}, which strongly suggests the origin from magnetic
scattering of the impurity states. The very tiny anisotropy in
underdoped Ba$_{1-x}$K$_x$Fe$_2$As$_2$ was thought to be ascribed to
the relatively small impurity potential as the dopant atom is
relatively far from the Fe-plane \cite{12}.

Very recently, it was surprisingly found in
Ba$_{1-x}$K$_x$Fe$_2$As$_2$ that, in contrast to the small positive
anisotropy of resistivity ($\rho_{\rm b} - \rho_{\rm a} > 0$)
existing at $x\leq$ 0.202, the sign of the anisotropy was reversed to
negative ($\rho_{\rm b} - \rho_{\rm a} < 0$) as $x=$0.235, which
begins at a temperature well above $T_{\rm N}$ \cite{24}. However,
as mentioned above, different mechanisms were considered to
interpret the anisotropy above and below $T_{\rm N}$, respectively.
As a consequence, there is a natural question about how the sign of
anisotropy evolves from totally positive to totally negative with
increasing the hole doping level. In this paper, we report on
anisotropic in-plane resistivity on detwinned hole-underdoped
Ca$_{1-x}$Na$_x$Fe$_2$As$_2$ crystals. $\rho_{\rm b} - \rho_{\rm a}
> 0$ is observed in the samples with $x$ = 0.11 and 0.14. In the
sample with $x$ = 0.19, however, an intersection happens at a
certain temperature of 110 K in $\rho_{\rm b} \sim T$ and $\rho_{\rm
a} \sim T$ curves; that is, $\rho_{\rm b} - \rho_{\rm a} < 0$ below
110 K, while $\rho_{\rm b} - \rho_{\rm a} > 0$ above 110 K and the
sign of resistivity anisotropy is the same as that observed in
the crystals with $x = $ 0.11 and 0.14. Subsequently, as $x$ is
increased further ($x\geq$0.24), the sign of this anisotropy can be
totally reversed, $\rho_{\rm b} - \rho_{\rm a} < 0$,  similar to the
results observed in Ba$_{1-x}$K$_x$Fe$_2$As$_2$ ($x$ = 0.235). Such
complicated evolution of anisotropy can be understood in terms of
combined effect of spin fluctuation and
the reconstructed Fermi surface (RFS) on anisotropy, with the effect of impurities included.

\section{EXPERIMENTAL DETAILS}

High-quality single crystals of Ca$_{1-x}$Na$_x$Fe$_2$As$_2$ were
grown by the self-flux method. The starting materials were CaAs,
NaAs, FeAs and Fe$_2$As, with the molar ratio of Ca/Na: Fe: As=1: 4:
4. The nominal compositions were $x$= 0.2, 0.23, 0.3, 0.35, and 0.4
respectively. After thoroughly grounding, the mixture was loaded into
an alumina crucible and then sealed in an iron crucible under 1.5
atm argon atmosphere. The reactants were heated to 1160 $\celsius$ in
a tube furnace protected with highly pure argon and kept at this
temperature for 10 h. Subsequently, the furnace was cooled down to
860 $\celsius$ at a rate of 5 $\celsius$/h. Finally the furnace was
cooled down to room temperature naturally by shutting off the power.
The actual chemical compositions were determined by energy
dispersive x-ray spectroscopy (EDS) to be 0.11, 0.14, 0.19, 0.24,
and 0.30 for the above five nominal compositions Na: Ca = 0.2-0.40,
respectively, with a standard instrument error of 10\%. Single-crystal x-ray diffraction (XRD) was performed on a SmartLab-9 diffracmeter(
Rikagu) from 10 to 65 $\deg$ with a scanning rate of
2$\deg$ per minute. The
crystals were cut in a rectangular shape along the tetragonal [110]
directions. The in-plane resistivity measurements were carried out
with the standard four-probe method by using a Quantum Design physical
property measurement system. The
in-plane resistivity along the orthorhombic \emph{a} and \emph{b} axes were measured
by a mechanical cantilever device similar to
Refs.\cite{8,11}.

\begin{figure}[t]
\centering
\includegraphics[width=0.48\textwidth]{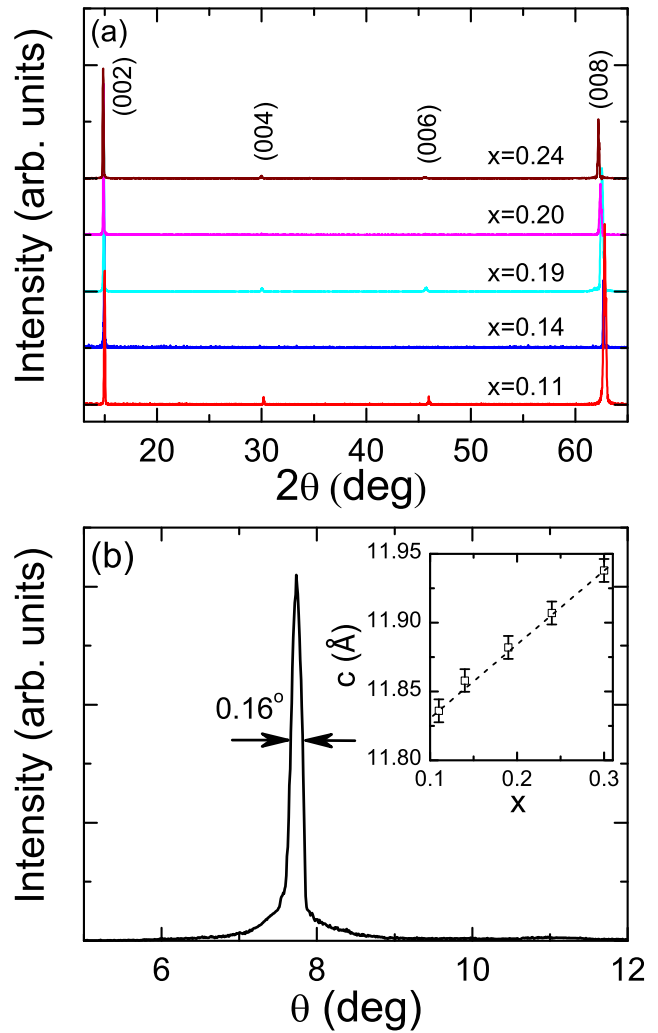}
\caption{(a) The single-crystal XRD patterns for Ca$_{1-x}$Na$_x$Fe$_2$As$_2$
single crystals with $x$ = 0.11$-$0.24. (b) Typical rocking curve of (002) reflection. The inset shows $x$ dependence of
the lattice parameter $c$ estimated from the data in (a).}
\end{figure}

\section{RESULTS AND DISCUSSION}

Figure 1(a) shows the single-crystal XRD patterns for the Ca$_{1-x}$Na$_x$Fe$_2$As$_2$
single crystals with $x$ = 0.11 $-$ 0.24. Only (0 0 2$l$) reflections show up, suggesting good orientation along the c axis for all the
crystals. The typical rocking curve of (0 0 2) reflection for the crystals is shown in Fig. 1(b). The full width at half maximum of the rocking curve is about 0.16$\deg$, indicating the high quality of the crystals. The inset of Fig. 1(b) shows lattice constant $c$ estimated from the data shown in Fig. 1(a), which increases nearly linearly with increasing Na doping level, consistent with the previous report on the polycrystalline samples of Ca$_{1-x}$Na$_x$Fe$_2$As$_2$ \cite{28}.

\begin{figure}[t]
\centering
\includegraphics[width=0.48\textwidth]{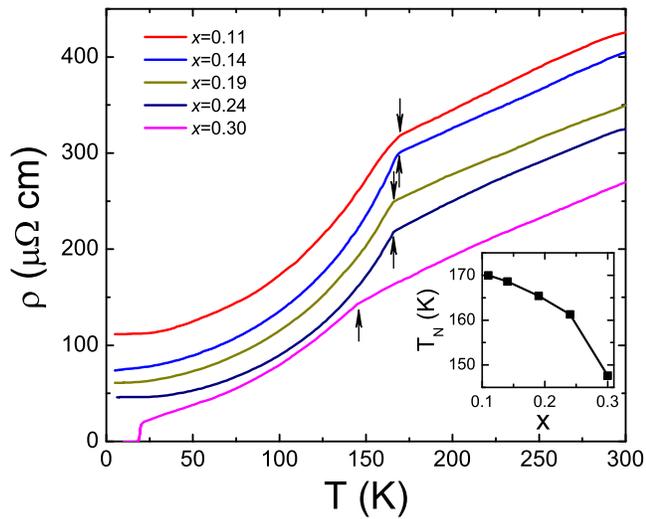}
\caption{Temperature dependence of the in-plane resistivity measured
on twinned Ca$_{1-x}$Na$_x$Fe$_2$As$_2$ crystals with $x$
=0.11-0.30. The arrows indicate the magnetic transition
temperatures, where the maximum of the derivative of resistivity
locates. The doping dependence of AFM transition temperature $T_{\rm
N}$ is shown in the inset.}
\end{figure}

The temperature dependence of the in-plane resistivity measured on
twinned Ca$_{1-x}$Na$_x$Fe$_2$As$_2$ crystals ($x$=0.11-0.30) is
shown in Fig. 2. The residual resistivity decreases with
increasing doping level and becomes less than 44 $\mu\Omega$ cm for
$x$ = 0.24, which is close to that in Ba$_{1-x}$K$_x$Fe$_2$As$_2$
crystal with 16\% doping level \cite{12}. A clear anomaly can be
observed for all of these samples in the temperature region of 148 -
170 K. Such single kink of anomaly in resistivity indicates that the
AFM and structural transitions take place at the same
temperature, the same as that observed in the 
Ba$_{1-x}$K$_x$Fe$_2$As$_2$ system \cite{27}. Therefore, we denote
the temperature where this anomaly locates simply as $T_{\rm N}$. The observed $T_{\rm
N}$'s are much higher than those observed in polycrystalline sample
for the each same doping level\cite{28}. $T_{\rm N}$ is plotted
against doping level in the inset of Fig. 2. With increasing the
doping level, $T_{\rm N}$ decreases quite slowly as $x\leq$ 0.24 and
then steeply as $x>$ 0.24. No superconductivity can be observed above 5
K as $x\leq$0.24 and the sample with $x$ = 0.30 shows the
superconducting transition at 20 K. The nonsuperconducting
underdoped region in Ca$_{1-x}$Na$_x$Fe$_2$As$_2$ is much wider than
that of Ba$_{1-x}$K$_x$Fe$_2$As$_2$ \cite{27}, in which
superconductivity emerges as $x>$0.14.

\begin{figure}[t]
\centering
\includegraphics[width=0.48\textwidth]{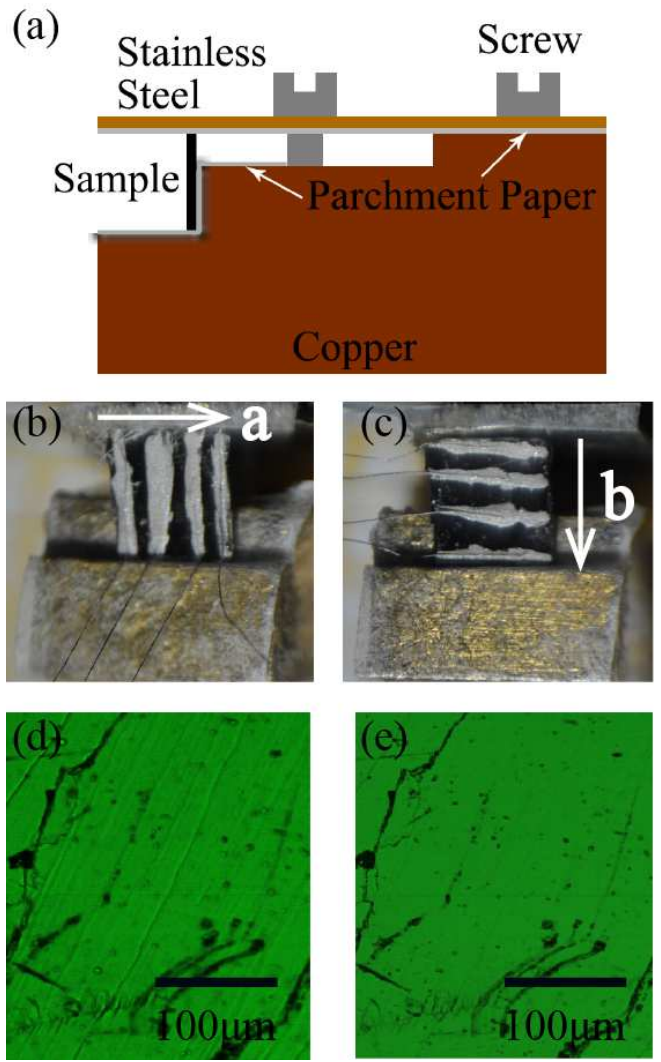}
\caption{(a) The schematic view of the setup for detwinning in this work. The parchmentpaper is used for insulating the sample from copper substrate and the stainless steel cantilever. (b), (c) A single-crystal sample of  Ca$_{1-x}$Na$_{x}$Fe$_2$As$_2$, mounted on the detwinning setup with the contacts aligned parallel and perpendicular to the direction of the strain pressure. (d), (e)
The polarized-light microscopic views for the surface of twinned and
detwinned case taken at 77 K respectively.}
\end{figure}

The twin boundaries in the orthorhombic phase of the underdoped iron-pnictides \cite{Tanatar} hampers probing the in-plane anisotropy and two methods (application of uniaxial strain or tensile and imposing an in-plane magnetic field) have been adopted to detwin such crystals so that the orthorhombic \emph{a} and \emph{b} axes can be distinguished \cite{8,11,10, 13,21,MF1,MF2} In this work, uniaxial strain was used for detwinning the crystals. The setup adopted in the study of the in-plane resistivity anisotropy is schematically shown in Fig. 3(a), which has widely been used in the previous works \cite{8,11} for detwinning iron-pnictide crystals. The mechanical strain was produced by tightening the screw nearby the sample and applied along the tetragonal [110] direction (which would become the orthorhombic \emph{a} or \emph{b} axis in the orthorhombic phase), as mentioned above. The typical configurations of current and voltage contacts for measuring resistivity along the orthorhombic \emph{a} and \emph{b} axes are shown in Fig. 3(b) and 3(c), respectively. Figures 3(d) and 3(e) are the typical polarized-light microscopy of the surface of Ca$_{0.89}$Na$_{0.11}$Fe$_2$As$_2$ crystal at the temperature of 77
K before and after detwinning, respectively. After detwinning, twin domain walls can no longer be seen, as shown in Fig. 3(e).

\begin{figure}[t]
\centering
\includegraphics[width=0.48\textwidth]{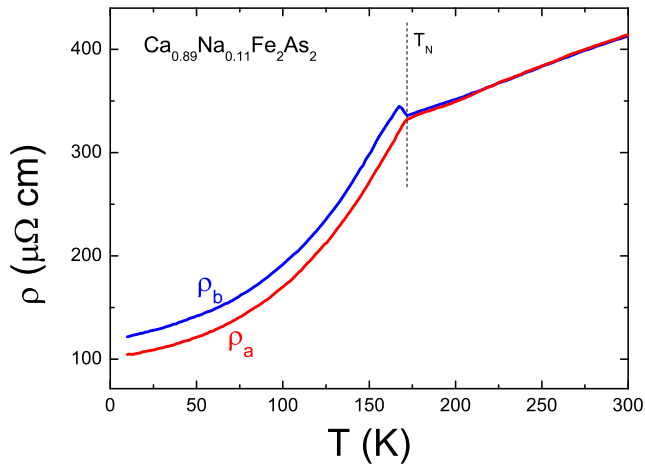}
\caption{Temperature dependence of the in-plane resistivity measured
on the detwinned Ca$_{0.89}$Na$_{0.11}$Fe$_2$As$_2$ crystal (blue,
$\rho_b$; red, $\rho_a$). $T_{\rm N}$ is determined from resistivity
measured on twinned crystal, as shown in Fig. 2.}
\end{figure}

Figure 4 shows the temperature dependence of the in-plane
resistivity along the \emph{a} and \emph{b} axes ($\rho_{\rm a}$ and $\rho_{\rm b}$)
of the detwinned Ca$_{0.89}$Na$_{0.11}$Fe$_2$As$_2$.
Small anisotropy of in-plane resistivity can be observed well above
$T_{\rm N}$ and it becomes larger as temperature is cooled close to
$T_{\rm N}$. A finite difference between $\rho_{\rm a}$ and $\rho_{\rm
b}$ in the AFM state remains to low temperature. The resistivity
along the \emph{b} axis is larger than that along the \emph{a} axis, e.g. $\rho_{\rm b}$
$>$ $\rho_{\rm a}$, which is similar to those observed in parent
CaFe$_2$As$_2$ and other electron-doped Ba- and Eu-122 crystals
\cite{8,9,10,11}.

\begin{figure}[t]
\centering
\includegraphics[width=0.48\textwidth]{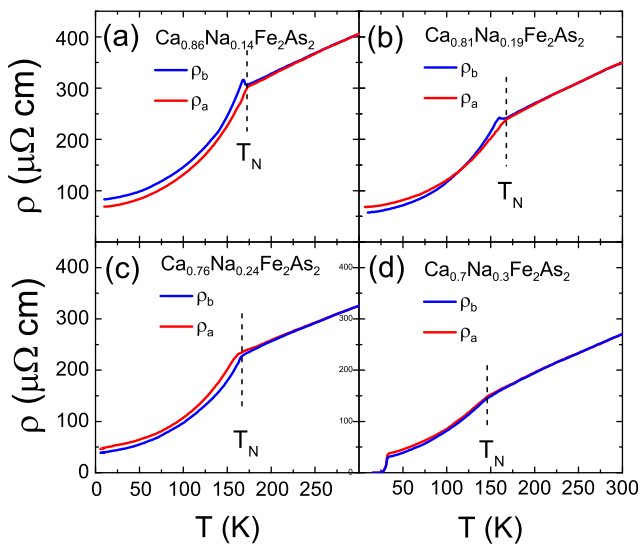}
\caption{Temperature dependence of the in-plane resistivity measured
on the detwinned Ca$_{1-x}$Na$_{x}$Fe$_2$As$_2$ crystal (blue,
$\rho_a$; red, $\rho_b$), with $x$ = 0. 14 - 0.24. }
\end{figure}

With the increase of the Na doping level, the anisotropy becomes small,
as observed for Ca$_{0.86}$Na$_{0.14}$Fe$_2$As$_2$ crystal [see Fig.
5(a)]. As shown in Fig. 5(c), $\rho_{\rm b}$ $<$ $\rho_{\rm a}$ can be
observed below a temperature well above $T_{\rm N}$ for the crystal
with $x$=0.24, which resembles previous observation in
Ba$_{1-x}$K$_x$Fe$_2$As$_2$ with nearly the same hole doping level
($x$ = 0.235) as ours \cite{24}. In a previous report for the
Ba$_{1-x}$K$_x$Fe$_2$As$_2$ \cite{24} system, the anisotropy changes
from $\rho_{\rm b}$ $>$ $\rho_{\rm a}$ for the sample with $x$ =
0.202 to $\rho_{\rm b}$ $<$ $\rho_{\rm a}$ for the sample with $x$ =
0.235. It should be noted that the sign reversal of anisotropy in
Ba$_{1-x}$K$_x$Fe$_2$As$_2$ happens in the superconducting samples.
While in the Ca$_{1-x}$Na$_x$Fe$_2$As$_2$ system, $\rho_{\rm b}$ $<$
$\rho_{\rm a}$ can already be observed in the nonsuperconducting
underdoped sample. Superconductivity emerges and a small difference
between $\rho_{\rm b}$ and $\rho_{\rm a}$ with $\rho_{\rm b}$ $<$
$\rho_{\rm a}$ can still be observed at low temperature for the
sample with $x > 0.3$, as shown in Fig. 5(d). The onset superconducting
transition temperature was enhanced by about 7 K from about 20 K at
ambient pressure after applying strain pressure, indicating that
superconductivity in Ca$_{1-x}$Na$_x$Fe$_2$As$_2$ is sensitive to
pressure. Although such uniaxial stress is usually low (typical
values of 5-10 MPa \cite{FisherR}), it has been reported that
superconductivity is induced by the uniaxial stress in underdoped
Ba(Fe$_{1-x}$Co$_x$)$_2$As$_2$ with $x$ = 0.016 and 0.025\cite{8}.
Such a dramatic enhancement of $T_{\rm c}$ under small uniaxial
stress applied within \emph{ab}-plane can be ascribed to the height of
anion from the Fe atom ($h$) in Ca$_{0.7}$Na$_{0.3}$Fe$_2$As$_2$ being
close to the optimal value 1.38 \AA \cite{h} (derived from the data
of polycrystalline sample \cite{28}, $h$ should be around 1.37 \AA
$~$as $x$ =0.3 in Ca$_{1-x}$Na$_x$Fe$_2$As$_2$). Uniaxial stress can
reduce the lattice parameters within \emph{ab}-plane, and consequently
results in an enhancement of $h$. As discussed in Ref.[28], $T_{\rm
c}$ can be enhanced sharply with increasing $h$ close to 1.38 \AA,
and small uniaxial stress can efficiently raise $T_{\rm c}$.

To clarify the nature of the sign-reversal of in-plane resistivity
anisotropy with increasing hole doping level in hole-doped
iron-pnictides, we studied the sample with the intermediate doping
level ($x$ = 0.19) to unveil how the sign of anisotropy develops
with increasing the hole doping level from  $x$ = 0.11 and 0.14 with
$\rho_{\rm b}$ $>$ $\rho_{\rm a}$ to $x$ = 0.24 and 0.30 with
$\rho_{\rm b}$ $<$ $\rho_{\rm a}$, as shown in Fig. 5(b). It is found
that sign reversal of the anisotropy happens in the special sample
with $x$ = 0.19 upon cooling, that is: $\rho_{\rm b}$ $>$ $\rho_{\rm a}$
starts to be observed at a temperature well above $T_{\rm N}$, and
$\rho_{\rm a}$ $>$ $\rho_{\rm b}$ occurs below about 110 K (much
less than $T_{\rm N}$). Therefore, sign reversal of in-plane
resistivity anisotropy occurs firstly at low temperature with
$\rho_{\rm b} > \rho_{\rm a}$ at high temperature, resulting
in an intersection between $\rho_{\rm a}$($T$) and $\rho_{\rm b}$($T$). These results suggest that there are competitive mechanisms
on the anisotropy in the AFM state.

\begin{figure}[t]
\centering
\includegraphics[width=0.45\textwidth]{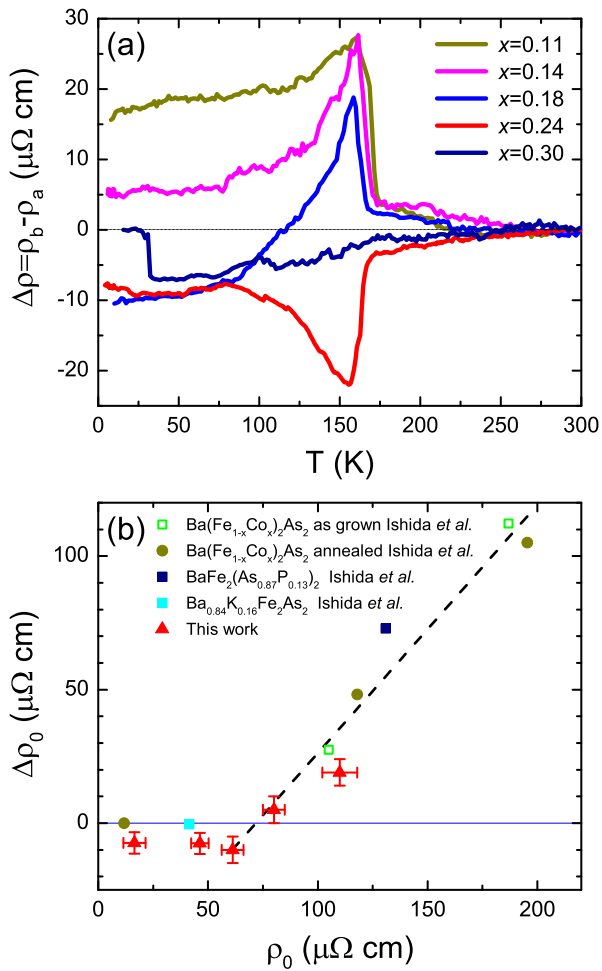}
\caption{(a) Temperature dependence of the in-plane resistivity
difference ($\Delta\rho = \rho_{\rm b}-\rho_{\rm a}$) for
Ca$_{1-x}$Na$_x$Fe$_2$As$_2$. (b) Difference in the residual
component of the in-plane resistivity at low temperature plotted
against the residual resistivity ($\rho_0$) obtained from
measurements on twinned crystals (as shown Fig. 2). The data other
than Ca$_{1-x}$Na$_x$Fe$_2$As$_2$ come from Ref. [6].}
\end{figure}

We plot the temperature dependence of the in-plane resistivity
difference $\Delta\rho=\rho_{\rm b}-\rho_{\rm a}$ for all the
samples in Fig. 6(a). The maximum of the magnitude of $\Delta\rho$
appears at temperatures a little below $T_{\rm N}$ ($T \approx
0.95T_{\rm N}$) for $x$ = 0.11 - 0.24. It is found that above
$T_{\rm N}$ there is already a finite resistivity anisotropy
($\Delta\rho\neq$0), suggesting the existence of nematic phase
\cite{25,34,35}. With increasing $x$, $\Delta\rho$ above $T_{\rm N}$
changes the sign from positive to negative at $x$ = 0.24. The sign
reversal of anisotropy occurs in the sample with $T_{\rm N}$ as high
as 162 K, which is much higher than $\sim$70 K in the
Ba$_{1-x}$K$_x$Fe$_2$As$_2$ case although their hole concentrations
are close to each other\cite{24}. The anisotropy of in-plane
resistivity above $T_{\rm N}$ has been theoretically ascribed to the
contribution from the anisotropic magnetic scattering due to the
spin (nematic) fluctuations associated with the
anisotropy electronic state (nematic state)\cite{25,37}, which has been proposed to interpret the sign
reversal in Ba$_{1-x}$K$_x$Fe$_2$As$_2$ above $T_{\rm
N}$ \cite{24,25}.

It was previously proposed that the mechanism responsible for the
anisotropy below $T_{\rm N}$ should be different from that for the
anisotropy above $T_{\rm N}$  due to the reconstruction of Fermi
surface induced by the AFM ordering\cite{9,26,24,FisherR}. It has
been theoretically suggested that the in-plane resistivity can be
larger along either the \emph{a} or \emph{b} direction, depending on the
shape of the Fermi surface, because the anisotropy of the Fermi
velocity (strongly connected to the morphology and topology of the
RFS) can lead to a large variation in the ratio of the Drude weight
along the two directions \cite{36}. However, recent theoretical work
by Sugimoto \emph{et al.} pointed out that the Drude weight gives
anisotropy opposite to experimental observation\cite{26}. It seems
to suggest that only the anisotropic RFS itself is not sufficient to
explain the observed in-plane resistivity anisotropy.

The impurity scattering within the FeAs layers has also been
proposed as one possible mechanism for the anisotropy of resistivity
below $T_{\rm N}$ in Co- and P-doped BaFe$_2$As$_2$ \cite{12,22,23}.
To illustrate the effect of impurity scattering on the anisotropy in
Ca$_{1-x}$Na$_x$Fe$_2$As$_2$, we plotted the difference in the
residual component of the in-plane resistivity ($\Delta\rho_0$) as a
function of the residual resistivity ($\rho_0$) at low temperature,
as shown in Fig. 6(b). The data from previous report on Co- and
P-doped BaFe$_2$As$_2$ are also included in Fig. 6(b) \cite{12}. As
$\Delta\rho_0 >$ 0 ($x = $ 0.11 and 0.14 in
Ca$_{1-x}$Na$_x$Fe$_2$As$_2$), a correlation between $\Delta\rho_0$
and $\rho_0$ is observed, that is, $\Delta\rho_0$ linearly decreases
with reducing $\rho_0$. Considering that the magnitude of $\rho_0$
reflects the level of impurity scattering, this correlation
indicates that the impurity scattering plays a significant role on the
anisotropy of the in-plane resistivity, as suggested by a previous
report \cite{12}.

As shown in Fig. 6(b), $\Delta\rho_0$ could reach zero as $\rho_0$ is
reduced to about 75 $\mu\Omega$ cm. Our group and Ishida \emph{et
al.} reported a negligible $\Delta\rho_0$ in
Ba$_{0.84}$K$_{0.16}$Fe$_2$As$_2$ which has $\rho_0$ ($\approx$ 41
$\mu\Omega$ cm) much less than 75 $\mu\Omega$ cm \cite{11,12}.
Ishida \emph{et al.} attributed this to rather weak impurity
potential \cite{12}.  However,  $\Delta\rho_0$ becomes negative in
Ca$_{1-x}$Na$_x$Fe$_2$As$_2$ when $\rho_0$ is reduced to less than
75 $\mu\Omega$ cm with increasing Na content. A similar behavior (
the finite negative $\Delta\rho_0$) has been reported in
Ba$_{0.765}$K$_{0.235}$Fe$_2$As$_2$ which has smaller $\rho_0$ than
 that in Ba$_{0.84}$K$_{0.16}$Fe$_2$As$_2$\cite{Yan}, Therefore, only
impurity scattering is not sufficient to understand the negative
$\Delta\rho_0$ in the regime with weak impurity level.

Very recently, Sugimoto \emph{et al.} tried to theoretically
reproduce the sign-reversal of in-plane resistivity anisotropy in
the AFM state based on the interplay of impurity scattering and the
anisotropic electronic states of RFS\cite{26}. The anisotropy was
thought to be dominated by the non-Dirac electron Fermi pockets near
the $\Gamma$ point in the existence of impurity potential \cite{26}.
In their model, negative $\Delta\rho_0$ can be realized as the
electron pockets disappear with increasing hole doping level
\cite{26}. However, there is no such experimental result for the
evolution of the RFS in the hole-underdoped samples up to now. As a
result, further experiments on these hole-underdoped crystals, such
as ARPES and quantum oscillation, are required to examine the
validity of the theoretical explanations.

No matter what is actually responsible for the sign-reversal of
$\Delta\rho$ deep inside the AFM state with increasing hole doping
level, the intersection between $\rho_a$($T$) and $\rho_b$($T$) in
Ca$_{0.81}$Na$_{0.19}$Fe$_2$As$_2$, with $\Delta\rho>$0 in a short
interval of temperature below $T_{\rm N}$ (110 K $<T<$ 165 K) while
$\Delta\rho<$0 deep inside the AFM state ($T<$ 110 K), cannot be simply
attributed to a single mechanism but should be the result of a
combined effect of different mechanisms. Apparently, $\Delta\rho$ in
the temperatures of 110 K $<T<$ 165 K inherits the positive sign of
$\Delta\rho$ above $T_{\rm N}$, suggesting that the sign of anisotropy in
this temperature region is still dominated by the magnetic
scattering of spin fluctuation although spin fluctuation becomes
weaker after entering the AFM state. Upon cooling from $T_{\rm N}$
to 110 K, the magnitude of $\Delta\rho$ continuously decreases in
the sample with $x$ = 0.19, indicating that the contributions from
different mechanisms lead to different signs of $\Delta\rho$ and
compete with each other below $T_{\rm N}$ in this doping level. However,
in the other doping levels we investigate, the data shown in Fig. 6(a)
suggest that the different mechanisms give the same signs of
contributions to $\Delta\rho$. In one word, the complicated
evolutions of resistivity anisotropy $\Delta\rho$ with hole doping
level and temperature shown in Fig. 6(a) suggest cooperative effect of
the contributions to $\Delta\rho$ from the different mechanisms:
spin fluctuation, impurity scattering, and anisotropic electronic
state of the RFS. These observations provide the hints to theoretical
explanation of the in-plane resistivity anisotropy in the AFM
state.

\section{CONCLUSION}

In conclusion, we investigated the in-plane resistivity anisotropy
in the detwinned hole-underdoped Ca$_{1-x}$Na$_{x}$Fe$_2$As$_2$
single crystals, and observed the sign reversal of in-plane
resistivity anisotropy with increasing hole doping level from $x$ =
0.11 and 0.14 to $x$ = 0.24 and 0.30 in the detwinned
hole-underdoped Ca$_{1-x}$Na$_{x}$Fe$_2$As$_2$ single crystals, and
an intersection between $\rho_{a}$($T$) and $\rho_b$($T$) deep
inside the AFM state for the crystal with $x$ =0.19.  These results
suggests that the anisotropic resistivity in the AFM state strongly
depends on the competition of the contributions from different
mechanisms. Such competition between the different mechanisms leads
to the complicated evolution of the anisotropy of the in-plane
resistivity with doping level.

\section*{ACKNOWLEDGEMENTS}

The authors would like to thank L. J. Zou, S. Y. Li, Z. Sun, Y. M.
Xiong and T. Wu for useful discussions. This work is supported by
the National Natural Science Foundation of China (Grants No.
11190021, No. 11174266, and No. 91122034), and the "Strategic Priority Research
Program (B)" of the Chinese Academy of Sciences (Grant No.
XDB04040100), the National Basic Research Program of China (973
Program, Grant No. 2011CBA00101), Anhui Provincial Natural Science
Foundation (Grant No. 1308085MA05), the Fundamental Research Funds
for the Central Universities (Programs No. WK2030020020 and No.
WK2340000035).

\end{document}